\documentstyle[amsmath,fleqn,twoside,espcrc2]{article}

\title{Modulus stabilization via non-minimal coupling}
\author{L.N. Granda\address{Department of Physics, Universidad del Valle\\
A.A. 25360, Cali, Colombia}}

\begin{document}%

\begin{abstract}
We propose a massless nonminimally coupled scalar field as a mechanism for 
stabilizing the size of the extradimension in the Randall-Sundrum I scenario.
Without needing to introduce self interactions terms we obtain a potential 
for the modulus field that sets the size of the fifth dimension. The minumum
of this potential yields appropriate values of the compactification scale
for small values of the coupling $\xi$.
\end{abstract}
\maketitle
\section{introduction}
An approach to the hierarchy problem has been proposed in
which large compactified extradimensions may provide an alternative
solution \cite{Arkani,Anton}. In these models the observed Planck mass 
$g_{p}$ is related to $M$, the fundamental scale of the theory, by
$M_{p}^2=M^{n+2}V_{n}$, where $V_{n}$ is the volume of the 
compactified dimensions. In $V_{n}$ is large enough, $M$ can be
of the order of the weak scale. This propierties rely on a 
factorizable geometry, namely the metric of the four familiar 
dimensions is independent of coordinates in the extradimensions. In
the works \cite{RSI,RSII}, Randall and Sundrum have proposed a higher
dimensional scenario which allow the existence of $4+n$ non-compact
dimensions in perfect compatibility with experimental gravity. In
this model the solution to the hierarchy problem rely on a non-
factorizable geometry, and consists (in the five dimensional case)
of a single $S^1/Z_{2}$ orbifold extradimension with two 3.branes of
opposite tension, residing at the orbifold fixed points \cite{RSI,RSII}.
Introducing a bulk cosmological constant and solving the Einstein's 
equations, gives the solution with non-factorizable geometry, that 
respects the four dimensional Poincare invariance
\begin{equation}\label{metric}
ds^2=e^{-2kr_{c}|\phi|}\eta_{\mu\nu}dx^\mu dx^\nu -r_{c}^2d\phi ^2
\end{equation}
where $-\pi \leq \phi \leq \pi $ is the extradimensional coordinate and 
$r_{c}$ is the interbrane distance, called compactification radius.This
solution holds only when the brane tensions and the bulk cosmological 
constant are related by the so called fine tuning conditions \cite{RSI}.
This condition ammounts to setting the four dimensional cosmological 
constant to zero, which is not a desired situation.\\
A similar scenario to the one propossed in \cite{RSI}, is that of 
Horava-Witten \cite{HW}, which arises within the context of 
supergravity nd M theory (see also \cite{Keha}for supergravity 
solutions).\\
Due to the exponential factor in the spacetime metric, a field confined
to the brane at $\phi =\pi $ with mass $m$, will have physical mass
$m e^{-kr_{c}\pi }$ and for $kr_{c}\approx 12$, the fundamental Planck 
scale $M$ is reduced to the weak scale (1Tev). However, as is well 
known, the dynamics does not determine the value of $r_{c}$, leaving
it a free parameter. A solution to this so called radion stabilization 
problem, has been found by adding a bulk scalar field which
generates the potential to stabilize the value of $r_{c}$. This
potential can appear from the presence of a bulk scalar field
with interaction terms that are localized to the two 3-branes 
\cite{WGold}. Alternarive stabilization mechanisms have been proposed
in \cite{Kim,Odin,Odin1,Gold}. The solution presented in the 
work \cite{Ben}, fixes the interbrane distance by adding to the
brane tension matter density and pressure. Depending on the kind
of matter, such solutions can be either stable or unstable under
small perturbations.\\
In this paper we present a solution to the
radion stabilization problem, by considering a massless bulk scalar
field, non-minimally coupled to 5-dimensional gravity. We find that,
without introducing self interaction terms on the branes, we 
obtain an effective potential, and the minima can be arranged to
yield a value of $kr_{c}\approx 12$, for small value of the bulk 
coupling constant $\xi$ . We present our solution in seccion II
and conclusions in section III.

\section{Effective potential for the coupled scalar field}

We denote the spacetime coordinates by $x^A=\{x^{\mu},y\}$,
where $x^{\mu}$, $\mu=0,...,3$ are Lorentz coordinates on the 3-branes
and $y=r_{c}\phi $, $-\pi \leq \phi \leq \pi $. The orbifold fixed 
points are at $\phi =0$ and $\phi =\pi $ and $r_{c}$ is the 
size of the extradimension.

The action for the bulk scalar field is given by

\begin{equation}\label{action}
S=\frac{1}{2} \int\!\! d^4 x\!\! \int\!\! d\phi {\sqrt {-{\mathcal G}}\,(G^{AB}\partial_{A}\Phi \partial_{B}\Phi 
- \xi R\Phi ^2}),
\end{equation}
where ${\mathcal G}=det[G_{AB}]$, $R$ is the bulk curvature for the metric (\ref{metric}) and is given by Eq. 
\begin{equation}
R=20\,\sigma '^2-8\,\sigma''
\end{equation}
where $\sigma '=\partial_{\phi }\sigma $. 
The $\phi$-dependent vacuum expectation value $\Phi (\phi )$ is 
determined by solving the equation of motion
\begin{equation}
-\frac{1}{r_{c}^2}\partial_{\phi }(e^{-4\sigma (\phi )}\partial_{\phi }\Phi)+
\xi \frac{e^{-4\sigma }}{r_{c}^2}(20\,\sigma '^2-8\,\sigma '')\Phi =0
\end{equation}
where $\sigma (\phi )=kr_{c}|\phi| $ and we take into account that according
to the solution of the Randall-Sundrum set up \cite{RSI},
$\sigma''=2kr_{c}(\delta (\phi )-\delta (\phi -\pi ))$. The solution to this 
equation excluding the fixed points $\phi =0,\pi $ is of the form \cite{WGold}
\begin{equation}
\Phi(\phi )=Ae^{(2+\nu)\sigma}+Be^{(2-\nu )\sigma },
\end{equation}
where $\nu =2\sqrt{1+5\xi }$. After replacing this solution into the scalar field 
action and integrating over $\phi$, we obtain an effective potential which 
depends on $r_{c}$ and $\xi$

\begin{eqnarray}\label{pot}
V(r_{c},\xi )=\frac{2A^2k}{\nu r_{c}}(2+\nu +10\xi +8\xi \nu )(e^{2\nu \pi kr_{c}}-1)\nonumber\\
+\frac{2B^2k}{\nu r_{c}}(2-\nu +10\xi -8\xi \nu )(1-e^{-2\nu \pi kr_{c}})
\end{eqnarray}
The coefficients $A$ and $B$ can be determined by imposing boundary conditions,
\begin{equation}\label{coef1}
A+B=\Phi (0)=V_{h}
\end{equation}
\begin{equation}\label{coef2}
Ae^{(2+\nu)\pi kr_{c}}+ Be^{(2-\nu)\pi kr_{c}}=\Phi (\pi )=V_{v}
\end{equation}
where the subindices $h$ and $v$ stand for hidden and visible branes respectibely. From Eqs. (\ref{coef1},\ref{coef2}) we 
obtain, after neglecting $e^{-\pi \nu kr_{c}}$ compared with 
$e^{\pi \nu kr_{c}}$, in the large $kr_{c}$ limit
\begin{align}
A=&V_{v}e^{-(2+\nu )\pi kr_{c}}-V_{h}e^{-2\nu \pi kr_{c}}\\
B=&V_{h}(1+e^{-2\nu \pi kr_{c}})-V_{v}e^{-(2+\nu )\pi kr_{c}}
\end{align}
Replacing the above given values of $A$ and $B$ into the Eq. (\ref{pot}) for the effective
potential, and neglecting terms of order $\xi ^2$, one obatins
\begin{multline}\label{effect}
V_{eff}(r_{c})=-(4+21\xi )[e^{-4\pi kr_{c}}\\
 +\left(\frac{V_{h}}{V_{v}}\right)^2 e^{-2\nu \pi kr_{c}}
 -2\frac{V_{h}}{V_{v}}e^{-(2+\nu)\pi kr_{c}}]\\
 +11\xi [e^{-2(2+\nu) \pi kr_{c}}\\
+\left(\frac{V_{h}}{V_{v}}\right)^2
 -2\frac{V_{h}}{V_{v}}e^{-(2+\nu)\pi kr_{c}}]
\end{multline}
where $V_{eff}(r_c)=V(r_c)/kV_{v}^2$ and we assumed $e^{-\nu\pi kr_{c}}\kern-0.7em\ll\kern-0.7em 1$. Considering $V_{h}/V_{v}=2$ 
this potential has a minimum at 
\begin{equation}
k\,r_{c}=\frac{1}{(\nu -2)\pi }ln(\nu ),
\end{equation}
with $\nu $ given by $\nu =2\sqrt{1+5\xi }$. Taking $\xi =1/265$ yields $kr_{c}\simeq 12$
which is the expected magnitude for $kr_{c}$ if we want to reduce the scale of quantum gravity
effects to the weak scale.
The possitive value of $d^2V(r_{c}) /dr_{c}^2$ signals the stability of the effective radion
potential. Here we have not considered the back reaction of the scalar field on the 
background geometry. This follows from the conditions $\xi \ll 1$ and $V_{v},V_{h}\ll M^{3/2}$, 
which allows the stress tensor for the scalar field to be neglected in comparison 
to the stress tensor induced by the bulk cosmological constant \cite{RSI,WGold}.

\section{Conclusions}
We have seen that a massless bulk scalar field, non minimally coupled to 5-dimensional
curvature, can generate a potential to stabilize $r_{c}$ for very small values of the
coupling constant $\xi $. Thus the large value of $k\,r_{c}$ arises not from small bulk
scalar mass \cite{WGold}, but from the small value of the coupling constant, and 
in addition, this stabilized modulus is obtained without introducing interaction 
terms on the branes.
If we consider quantum effects arising from bulk scalars, then non trivial vacuum 
energy appears. This bulk Casimir effect should play a remarkable role in the 
stabilization mechanism, as it gives contribution to both, the brane and the bulk
cosmological constants (see \cite{Odin} for works in this direction). It would be 
worthwhile to consider the effect of the bulk scalar fiel on the background geometry, and
also to explore the role of this field in cosmological models with accelerating expansion.

\section{Acknowledgments}
I would like to thank the organizers of the ¨Renormalization Group and Anomalies in Gravity and
Cosmology¨ meeting. This work has received financial support by 
COLCIENCIAS, contr. N° 1106-05-11508.

\end{document}